\documentstyle[psfig,12pt,aaspp4]{article}

\def\Ntot{84}
\def\Nopt{76}
\def\Nuse{70}
\def\kms{km~s$^{-1}$}
\def\arcmin{$^{\prime}$}
\def\arcsec{$^{\prime\prime}$}
\def\hal{H$\alpha$}
\def\be{\begin{equation}}
\def\ee{\end{equation}}
\def\about{$\sim$}
\def\etal{{\it et al.}}
\def\eg{$e.g.$}
\def\ie{$i.e.$}
\def\deg{$^{\circ}$}

\def\rd{${R_{\rm d}}^{\circ}$}
\def\rA{$R_{\rm A}$}
\def\ropt{$R_{\rm opt}$}
\def\vpec{$V_{\rm pec}$}
\def\epso{$\epsilon_{\rm obs}$}
\def\epst{$\epsilon_{\rm true}$}

\begin{document}
\hskip 3.5in{\hskip 10pt \today}
\title{Seeking the Local Convergence Depth. I. TF Observations of the Clusters
A168, A397, A569, A1139, A1228, and A1983}
 
\author {Daniel A. Dale\altaffilmark{1}, Riccardo Giovanelli, Martha P. Haynes, 
Marco Scodeggio\altaffilmark{2}}
\affil{Center for Radiophysics and Space Research
and National Astronomy and Ionosphere Center\altaffilmark{3}, Cornell University,
Ithaca, NY 14853}
\affil{dale,riccardo,haynes,scodeggi@astrosun.tn.cornell.edu}

\author {Eduardo Hardy}
\affil{Department of Physics, Laval University, Ste--Foy, P.Q., Canada G1K 7P4}
\affil{hardy@phy.ulaval.ca}

\author {Luis E. Campusano}
\affil{Observatorio Astron\'{o}mico Cerro Cal\'{a}n, Departamento de 
Astronom\'{\i}a, Universidad de Chile, Casilla 36-D, Santiago, Chile}
\affil{lcampusa@das.uchile.cl}

\altaffiltext{1}{Department of Physics, Cornell University}

\altaffiltext{2}{Now at European Southern Observatory, Karl Schwarzschild Str. 
2, D--85748 Garching b. M\"unchen, Germany}

\altaffiltext{3}{The National Astronomy and Ionosphere Center is operated by 
Cornell University under a cooperative agreement with the National Science 
Foundation.}

\begin{abstract}

We present first results of an all-sky observing program designed to 
improve the quality of the I band Tully-Fisher (TF) template and to obtain the
reflex motion of the Local Group with respect to clusters to $z$ \about\ 0.06.  
We are obtaining between 5 and 15 TF measurements per cluster on a sample of 
\about\ 50 clusters at intermediate redshifts (0.02 $\lesssim z \lesssim$ 0.06).  Presentation of the data for seven Abell clusters of galaxies is given here.  
This data incorporates methods for estimating the true inclination of a 
spiral disk, an observational parameter undervalued for small angular-sized 
galaxies or for galaxies observed in poor seeing conditions.
\end{abstract}


\section {Introduction}
Claims of large amplitude, coherent flows on scales up to 100$h^{-1}$ Mpc 
(Lauer \& Postman 1994:LP; Courteau \etal\ 1993) are difficult to accomodate 
with current cosmological scenarios (Gramann \etal\ 1995).  Observationally, 
the LP claim has been challenged by several groups (\eg\ Riess \etal\ 1995 and
Giovanelli \etal\ 1996:G96), but a clear--cut solution of the disagreement is
still wanting, because the sample of supernova type Ia of Riess \etal\ (1995) is 
too sparse (Watkins \& Feldman 1995) and the Tully--Fisher (TF) sample of G96 is of limited depth.  The G96 sample of peculiar velocities (\vpec's) obtained 
using the TF method is referred to a template relation based on 555 galaxies in 
24 clusters to $cz\simeq 9000$ \kms\ (Giovanelli \etal\ 1997a,b:G97a,b).  If the 
\vpec\ field is characterized by strong spatial gradients, a shallow sample 
would not provide a quantitatively effective test of large--scale flows.  This is particularly important in view of the claims (Scaramella \etal\ 1994, 
Tini--Brunozzi \etal\ 1995, and Branchini \etal\ 1996) that asymptotic 
convergence of the Local Group reflex motion may only be reached at distances in 
excess of 10,000 \kms.  This is the first of a series of papers that aim at the 
direct determination of the TF relation for an all--sky cluster set extending to 
distances exceeding the highest reported values of the convergence depth.

A second issue of concern regards the amplitude of systematic errors in the TF
template relation used to measure \vpec.  Though the proximity of the G97a,b 
sample provides the stretch in galactic properties necessary to determine 
accurately the TF slope, it conversely works against pinning down the relation's 
zero point offset; peculiar velocities introduce relative distortions of 
redshifts that are larger for nearby galaxies than for distant ones.  Thus, the
benefits of measuring peculiar velocities at higher redshifts are twofold: the
contribution to the Local Group velocity field arising at large scales is better
understood {\it and} the (offset) accuracy of all \vpec\ measurements 
improves.

We are complementing the sample of G97a,b by obtaining between 5 and 15 TF 
measurements in each of approximately 50 clusters with 6000 \kms\ $\lesssim cz 
\lesssim$ 18,000 \kms.  Explicitly, the goals for this project are to obtain (a) 
a high quality template TF relation to which local motions can be accurately 
referred, and (b) a bulk flow dipole solution for a volume of radius 18,000 \kms.

In order to make the data publicly available on the shortest possible time 
scale we shall present partial results of our survey, as we complete data 
analysis for sizable fractions of the cluster set.  In this paper we present the 
project's first data installment:  TF measurements for \Ntot\ galaxies in the
fields of six Abell clusters, one of which, A1983, we actually treat as two 
separate clusters (see section 6).  In addition, we discuss and illustrate with 
examples some of the systematic effects involved in deriving true TF templates.  
The following section details the selection of clusters and galaxies for this 
project.  We describe the imaging of the clusters in section 3 and the optical 
spectroscopy of the galaxies in section 4.  In Section 5 we review the method 
used to estimate actual disk inclinations from the observed, seeing-distorted 
inclinations.  Section 6 presents the relevant TF data.

\section{Selection of Clusters and Galaxies}		

Clusters are selected using the Abell \etal\ (1989, hereinafter ACO) rich cluster catalog.  Details on the selection of the cluster sample will be provided in a 
separate paper.  To determine locations of target fields to be imaged, we 
visually scanned the Palomar Observatory Sky Survey plates for regions in the 
clusters containing promising disk systems appropriate for TF work.    

The selection of target galaxies for this study stems directly from the images
obtained.  Each image in clusters chosen for spectroscopy is searched for spiral disks with the following properties:  inclinations $\gtrsim$ 40$^{\circ}$; lack 
of dominating bulges -- bulgy disk systems tend to be gas deficient and thus 
undesirable for spectroscopy; no apparent warps/interacting neighbors; and no 
nearby bright stars which may affect flux measurements.  Coordinates and position angles for these objects are obtained from the Digitized Sky Survey\footnote{The 
Digitized Sky Surveys were produced at the Space Telescope Institute under U.S. Government grant NAG W-2166.  The images of these surveys are based on 
photographic data obtained using the Oschin Schmidt Telescope on Palomar Mountain and the UK Schmidt Telescope.} and are accurate to $<$ 2\arcsec.

Finally, we have earmarked at least two clusters in each semester to be observed
from both hemispheres.  In this way any systematic differences between data taken
in the North and South and between data taken from season to season can be
quantified.  Preliminary results indicate negligible differences; a detailed 
analysis will be presented when the bulk of the data has been obtained.

\section {Imaging}	     				

CCD images of the clusters were obtained at the KPNO and CTIO 0.9 m telescopes.  
This installment only contains KPNO data observed during the nights of October 
5--16 1994, September 14--21 1995, September 24--October 1 1995, and February 
13--19 1996.  Images were taken in the I band to match the observations of the 
nearby study of G97a,b and for the reasons outlined therein.   The $2048^2$ T2KA 
CCD (0.68\arcsec\ pixel$^{-1}$) used at the KPNO 0.9 m allows 23\arcmin\ fields.  Besides being useful in mapping out cluster fields efficiently, such wide fields 
also help to provide sufficient area in which to determine the sky background.  A
600 second integration time was typically used for each exposure; integrating for
ten minutes probes the low surface brightness edges (\about\ 24.5 mag 
arcsec$^{-2}$) of galaxies adequately and is generally not long enough to produce saturated galactic cores.

\subsection{Photometry Data Reduction}

We have made an effort to employ a method of data reduction consistent with the 
G97 reductions, which will be discussed in Haynes \etal, in preparation.  Here 
we summarize those details.  Both standard and customized IRAF\footnote{IRAF 
(Image Reduction and Analysis Facility) is distributed by the National Optical 
Astronomy Observatories, which are operated by the Association of Universities 
for Research in Astronomy, Inc. (AURA), under a cooperative agreement with the 
National Science Foundation.} packages are used.  Among the most crucial 
reduction tasks is obtaining adequately ``flat'' images due to the increased sky 
fluctuations present in the infrared, relative to the optical bands.  As such, we create flat--field templates by median--combining at least 30 frames that are 
free from relatively bright stars and relatively large galaxies.  Once 
flat--fielded, each galaxy frame is background corrected by subtracting off the 
mean value of several sky regions near the galaxy that are free from stars or 
significant cosmic ray hits (this is done separately for each galaxy).  The 
uncertainty in the sky background is less than 0.2\%.  Moreover, cosmic rays and 
also any stars near the galaxy are masked to prevent distortions of the ensuing 
isophotal ellipse fitting, a procedure that interpolates over masked regions.
  
The procedure of ellipse fitting employs the GALPHOT surface photometry package 
written for IRAF/STSDAS\footnote{STSDAS (Space Telescope Science Data Analysis 
System) is distributed by the Space Telescope Science Institute, which is 
operated by AURA, under contract to the National Aeronautics and Space 
Administration.} by W. Freudling, J. Salzer, and M.P. Haynes.  We fit ellipses to increasing galactic radii until the background is reached.  The bulk of the 
instrumental magnitude is determined from data within the outermost isophote, but a small yet non--zero contribution, found by extrapolating the ``curve of 
growth'' of the surface magnitude of each spiral galaxy to infinity (assuming the light drops like that of an exponential disk), is also included (see G94 for 
details on this procedure) to infer a ``total'' magnitude, m$_{\infty}$.  This 
contribution is typically a few hundredths of a magnitude.  In this paper we also report the values of the isophotal magnitude, $m_{\rm 23.5}$, measured at the 
23.5 mag arcsec$^{-2}$ isophote.  Errors on $m_{\rm 23.5}$ and $m_{\infty}$ 
typically hover between 0.02 and 0.04 mag -- the corrections for internal 
extinction introduce further uncertainty, as discussed in the next section.

Instrumental magnitudes are converted to apparent magnitudes using Landolt 
(1992) standard stars.  Many observations of standards were made each night in 
the R and I bands to cover a broad stretch in both airmass and R-I color.  We 
find a small dependence on airmass (\about\ 0.1 mag airmass$^{-1}$) and a 
negligible color effect.  This is fortunate since we only have I band images; 
assuming an R-I galactic color of 0.5 mag yields negligible color corrections.  
Finally, we emphasize that only data taken on nights where the photometric zero 
point is determined to 0.02 mag or better are used.

The ellipse fitting is also important in determining the ellipticity of the 
disk and hence inferring its inclination.  We discuss in detail the process of 
determining ``true'' disk inclinations in section 5.

\subsection {Corrections to I Band Magnitudes}

We use Burstein and Heiles' (1978) method for tabulating galactic extinction values by averaging entries in the {\it Third Reference Catalogue of Bright
Galaxies} (de Vaucouleurs \etal\ 1991) found near the cluster centers.  We 
convert B-Band extinctions to I band via $A_I$ = 0.45$A_B$; values of $A_I$
for the seven clusters presented here range from 0.00 to 0.18 mag.  The internal extinction correction is applied using the procedure outlined in G97a, whereby 
the decrease in apparent luminosity of late-type galaxies due to their 
dusty disks can be quantified as
\be
 \Delta m_{\rm int} = - f(T) \; \gamma(W_{\rm cor}) \; log(1-e)
\ee
where $\gamma$ ($\lesssim$ 1.0) depends on the corrected velocity width 
$W_{\rm cor}$ (section 4.3)
and $e$ is the ellipticity of the spiral disk (the correction $\Delta m_{\rm int}$ 
is slightly smaller for early, less dusty gals:  $f(T)$=0.85 for types $T$ 
earlier than Sbc; $f(T)$=1 otherwise).  In other words, the internal extinction 
depends on the size of the system and on how inclined the disk is to our 
line of sight.  An error for $\gamma$ of 0.15$\gamma$ is assumed.  We apply a 
cosmological k--correction: $k_I = (0.5876-0.1658T)z$ (Han 1992).  Finally, we 
are careful 
to apply a small correction, $\Delta m_T$, for the TF dependence on 
morphological type found in G97b.  Thus, the apparent magnitudes given in this 
paper are
\be
m_{\rm cor} = m_{\infty} - A_I - k_I - \Delta m_{\rm int} - \Delta m_T.
\ee
Note that $A_I$, $\Delta m_{\rm int}$, and $\Delta m_T$ are all positive 
quantities in equation (2).

\section {Spectroscopy}				

Though the 21 cm spectral line is a frequently adopted tool in TF work --- the HI distribution in galaxies typically extends further out than that of HII regions 
and thus samples rotation curves at larger radii --- the characteristics of our 
sample favor the use of optical spectroscopy.  Too few radio telescopes exist 
with the sufficient sensitivity, spatial and velocity resolution, and bandwidth 
necessary for an all--sky survey of intermediate distance clusters.  The dishes 
at Nan\c cay and Arecibo would be useful, but cannot cover the whole sky.  In
cluster fields at $cz \gtrsim$ 10$^4$ \kms, confusion becomes a 
problem and, especially at Arecibo, the interference environment makes
observations more difficult.  Array instruments still have limited correlator
power to cover the large bandwidth required for cluster observations with 
sufficient resolution.  Furthermore, many cluster spirals are HI deficient 
(Giovanelli \& Haynes 1985).  For these reasons we have chosen to adopt emission 
line spectroscopy near the \hal\ line to measure velocity widths; we detected 
\hal\ emission in 91 out of 105 targeted galaxies observed in these seven 
clusters.  The spectroscopy was carried out at the Palomar Observatory 5 m 
telescope over the course of four observing runs.  The nights of observing 
included:  October 22--23 1995, December 13--18 1995, Jan 24--25 1996, and May 
14--15 1996.  

For this project we used the red camera (5200 -- 11,000 \AA) of the 
Double Spectrograph (Oke and Gunn 1982) at the Cassegrain focus ($f$/15.7).  
The combination of the 1200 lines mm$^{-1}$ grating and a 2\arcsec\ wide slit 
yields a dispersion of 0.65 \AA\ pixel$^{-1}$ and a spectral resolution of 1.7 
\AA\ (equivalent to 75 \kms\ at 6800 \AA).  Along the cross--dispersion axis, the spatial scale is 0.46\arcsec\ pixel$^{-1}$.  Since we are interested in observing cluster members with redshift velocities of up to \about\ 25,000 \kms, the angle 
of the grating is set in order to observe the \hal\ line at wavelengths anywhere 
between 6560 \AA\ (\about\ rest wavelength) and 7230 \AA\ ($cz$ \about\ 30,000 
\kms).

As an observing strategy, we perform a five minute test integration on each
spectroscopic target.  In this way we are able to determine ``on the fly'' the
requisite exposure time to adequately sample the outer disk regions.
Furthermore, this test exposure determines if the galaxy is even useful to our
work; the galaxy may lie in the foreground or background of the cluster or it may
contain little or no \hal\ emission.  If the observation is deemed useful, a 
second integration typically ranges between 15 and 45 minutes.

\subsection{Spectroscopy Data Reduction}

Standard and customized IRAF tasks are used in reducing the spectroscopic data.
The spectra are wavelength calibrated using both night sky lines and HeNeAr 
comparison lamps.  The residual uncertainties in the spectral calibration 
translate to a typical velocity uncertainty of \about\ 7 \kms, a figure which 
is added in quadrature to the errors from the rotation curve fitting (below) in 
determining the total error for each data point.  The night skylines and the 
galactic continuum are fit with polynomials and then subtracted from the spectra.  Finally, the \hal\ rotation curve is computed.  In \about\ 1\% of the cases the 
emission of a N[II] line (6584 \AA) is stronger and extends to larger radii than 
that of the \hal\ emission; in these cases the N[II] curve is used.  At each 
position along the cross--dispersion axis we fit a Gaussian profile to the light 
distribution (averaged over three pixels in the spatial direction).  The position of the Gaussian peak is taken to be the line centroid for that spatial location.  Since the seeing is typically 1\arcsec\ to 2\arcsec, the rotation curves are 
oversampled by a factor of two to three.  The galactic center-of-light is defined
to be at the peak of the galaxy continuum light profile formed by summing the 
data along the dispersion direction.

\subsection{Calculating Systemic Velocities and Velocity Widths}

It is necessary to extract two parameters from each rotation curve:  the systemic velocity (\ie\ redshift) and the Doppler velocity width.  In an idealized galaxy, where both arms of the rotation curve are smoothly traced out beyond the turnover radius, the two velocity extrema are all that are needed -- the average of the 
extrema is the systemic velocity and the difference is the velocity width.  
Unfortunately, rotation curves can exhibit erratic behavior.  To determine 
velocity widths, we sort the velocity data points and calculate the difference 
between the 10\% and 90\% velocities (An N\% velocity is greater than N\% of all 
the velocity data points).  This technique is effectively a ``pixel histogram,'' 
similar in flavor to the portrayal of raw 21 cm data.

The systemic and rotational velocity extracted from an observed rotation curve 
depends on whether a photometric or kinematic center of symmetry is assumed.  In 
the first case, the systemic velocity is assigned the value at the location where the galaxy light peaks.  In the upper panel of figure 1, the rotation curve is 
folded about such a photometric center, exhibiting asymmetry in the outer parts. 
Alternatively, by forcing a match between the asymptotic rotational velocity in 
the outer parts of the galaxy, a ``kinematical centering'' can be obtained.  This kinematic center is merely the average of the 10\% and 90\% velocitites.  In the 
lower panel of figure 1, the rotation curve of the same galaxy in the upper panel is folded about the kinematic center, which is shifted from the photometric 
center by 0.92\arcsec.  The systemic velocity differs by 34 \kms\ between the two cases.  In what follows, we will adopt the kinematical centering technique.

\subsection{Corrections to Velocity Widths}

Besides avoiding the spatial confusion problem encountered by single dish radio
telescopes, important in observing galaxies clustered together at intermediate 
redshifts, optical emission line spectroscopy provides detailed mapping of the 
rotation curve along the chosen position angle.  This mapping facilitates 
estimating the impact of the optical rotation curve (ORC) shape (\eg\ still 
rising at the optical radius) on the determination of the velocity width.  On the other hand, advantages of using the 21 cm line include that 1), velocity widths 
are ``position angle--independent'' and 2), HI typically extends farther than the optical radius; an HI velocity width is more likely to have included light near 
or beyond the turnover radius, the point at which the rotation curve begins to 
fall.  Hence, in many instances, 21 cm work gives more accurate velocity widths 
than could be measured optically.  Haynes \etal\ (1997) reports 21 cm 
measurements for \about\ 840 galaxies at redshifts of up to $z$ \about\ 0.04, a 
data set which has a small overlap with our sample; we forego \hal\ and use 21 cm widths when they are available (see Table 2).

The first correction applied to the observed velocity width, $W_{\rm obs}$, is an additive term ${\Delta}_{\rm sh}$ that depends on the radial extent and shape of 
the ORC.  Small corrections are applied to the velocity widths in cases when the 
ORC does not extend to \ropt, the distance along the major axis to the isophote 
containing 83\% of the I band flux.  This is reported by Persic \& Salucci (1991 
\& 1995) to be a useful radius at which to measure the velocity width of ORCs.  The value 
of ${\Delta}_{\rm sh}$ is larger for non-asymptotic ORCs, but is usually less 
than 10\% of the velocity width; it is estimated by fitting a ``universal 
rotation curve'' function to the data as advocated by Persic \& Salucci, and
extrapolating it to \ropt.  We remark that this application of the Persic \&
Salucci formalism is merely a tool for a mild extrapolation of the outer 
part of the rotation curve for the estimation of uncertainties, and is not a 
test of the
relation's universality.  It should be noted that we do not use galaxies whose 
rotation curves are still rising significantly at the end of the \hal\ 
distribution (\eg\ curves exhibiting only solid--body rotation are deemed
unusable for TF work).  The other corrections to the observed velocity width are 
the factors 1/(1+$z$) and 1/sin$i$, the first of which accounts for the
cosmological broadening of rotation curves and the second corrects for the 
inclination $i$ at which a given disk is observed.  The corrected optical 
velocity width can be expressed as
\be 
W_{\rm cor} = {{W_{\rm obs} + {\Delta}_{\rm sh}} \over {(1+z)\sin i}}.
\ee

\subsection{Errors in Velocity Widths}

Many factors contribute to the uncertainty in a velocity width measurement,
including the inclination $i$.  In practice we measure a disk's ellipticity,
$\epsilon$, so to describe inclinations in terms of ellipticities, we follow 
Hubble (1926) in assuming spiral disks are oblate spheroids with intrinsic axial
ratio $q$:
\be
 \cos^2i = {{(1-\epsilon)^2 - q^2} \over {1 - q^2}}.
\ee
We also follow the convention outlined in Giovanelli \etal\ (1994;G94) in 
adopting a value of $q$=0.13 for type Sbc and later and a value of $q$=0.2 for 
morphological types earlier than Sbc.  Formally, the error in the inclination 
$i$ can be expressed as
\be
\sigma_i^2 = \Bigl({di \over d\epsilon}\Bigr)^2 \sigma_\epsilon^2 + 
             \Bigl({di \over dq       }\Bigr)^2 \sigma_q^2
\ee
where $\sigma_\epsilon$ is the uncertainty in the disk ellipticity measurement 
and $\sigma_q$ is the error in $q$, assumed to be 0.15$q$.

We can now quantify the error in the velocity width:
\be
\sigma^2_w = {\sigma^2_{w,{\rm obs}} + \sigma^2_{\rm sh}\over {(1+z)^2\sin^2 i}} + 
\sigma^2_i W^2_{\rm cor}\cot^2 i
\ee
where $\sigma_{w,{\rm obs}}$ is the error in the observed velocity width and
$\sigma_{\rm sh}$ is the error in the shape correction applied, typically 
0.25$\Delta_{\rm sh}$.  In the few cases where a 21 cm width is used, we follow 
G97a in computing errors in the widths.

\section {Inclination Corrections} 

Knowledge of the true rotational velocity widths of galaxies is crucial to TF 
studies, yet many circumstances combine to yield incorrect measurements.  Among 
these circumstances in optical spectroscopy work are misalignment of the slit 
with the major axis of the galaxies and insufficient integration times that 
prevent unambiguous measurements in the outer parts of the ORC.  The largest 
error contributed to estimates of the velocity width in general derives from 
uncertainty in the angle of inclination, $i$, of the disk to the line of sight.  
As described in the previous section, this geometric effect can be easily 
accounted for by dividing the observed velocity width by sin$i$.  In addition,
a knowledge of the correct ellipticity of a disk is important in the derivation
of the internal extinction correction to the total magnitude, as discussed in 
section 3.2.  It is the ability to correctly measure the ellipticity that we are 
concerned with here.

It is well--known that ground-based observations are affected by the atmospheric 
seeing conditions.  Saglia \etal\ (1993;S93) have recently provided a thorough 
analysis concerning the effects of seeing on spheroidal systems.  The turbulent 
atmosphere increasingly affects the observed ellipticities, \epso, as the 
angular size of galaxies decreases.  Here we consider that effect on spiral 
disks.    

\subsection{Measuring True Ellipticities}

In order to gauge the relationship between \epso\ and the undistorted 
ellipticity \epst, we have taken a sample of eleven large angular--sized galaxies (major axis $>$ 5\arcmin) with a broad range of inclinations and bulge--to--disk 
ratios and blurred their (I band) images by convolving them with a point 
spread function (psf) of varying width.  The galaxies used in this exercise are a
mix of Sa and Sc type galaxies and had \epst\ ranging between 0.311 and 0.865 and
bulge--to--disk ratios between 0 and 2.  We assume that the unconvolved images 
yield \epst. 

It is necessary in this exercise to mimic the actual seeing profile as closely 
as possible.  Though the seeing quality for an observation is typically 
conveyed by measuring the FWHM of a Gaussian fit to a stellar profile, it 
is known that stellar profiles do not drop as steeply outside of their cores as 
Gaussians do (Woolf 1982).  To simulate this effect we have followed the method 
of S93 which employs atmospheric turbulence theory and empirical results to 
predict the stellar psf and its parameters.  Letting the stellar psf be 
$p_{\gamma}(r)$, it can be shown that
\be
 p_{\gamma}(r) = {1 \over {2\pi}} {\int_0}^{\infty} dk \; k \;  
e^{-(kb)^{\gamma}} J_0(kr), \; \; \; \; \gamma = 5/3
\ee
where r is the radial distance, $J_0$ is the standard Bessel function, $b$ is 
FWHM/2.921, and $\gamma$ is a parameter predicted to be 5/3 by theory.

For a fixed \epst, we expect \epst/\epso\ to increase when either the size of 
the seeing disk increases or the angular size of the galaxy decreases.  We have 
thus parametrized the change in ellipticity as a function of the ratio between 
the measured seeing FWHM and the face--on disk scale length, \rd\ (see G94 for 
details on measuring \rd).  We thus define
\be
 \eta \equiv {{\rm FWHM} \over {R_d}^{\circ}}.
\ee
We have convolved each galaxy image with a range of values for $\eta$ and 
measured the resulting ellipticities.  Figure 2 gives our results for the eleven 
galaxies used in this study, plotting \epst/\epso\ as a function of $\eta$.  
We find that, within the uncertainties of our simulations, the change in the ratio \epst/\epso\ is independent of disk inclination and bulge--to--disk ratio. 
Fitting a linear relation to the data gives us a general method for estimating \epst\ as a function of the observable quantities \epso, FWHM, and \rd. 
We see that the change in ellipticity is not severe; there is only a \about\ 10\% 
difference even for $\eta$ \about\ 1, the extreme case when the seeing is comparable to the disk scale length.  More exactly,
\be
 \epsilon_{true} = \epsilon_{obs} + 0.118 \; \eta \; \epsilon_{obs}
\ee
where the 1-$\sigma$ errors in the offset and slope are 0.004 and 0.011, 
respectively.

\subsection{Examples of the Impact of Inclination Corrections}

As mentioned previously, both the velocity width and the total magnitude depend 
on the disk inclination.  Quantitative examples of the corrections applied are 
useful to get a feeling as to how TF parameters are changed.  Suppose two Sc 
galaxies with total velocity widths of 300 \kms\ are observed, with one inclined 
to the line of sight at 85.0\deg\ (galaxy `A') and the other at an inclination 
of 48.0\deg\ (galaxy `B').  This means the projected velocity widths would be 
299 \kms\ and 223 \kms\ respectively for galaxies A and B (ignoring cosmological 
effects).  Furthermore, suppose the sizes of the galaxies and the seeing 
conditions yield $\eta$ = 1.00 in each case (an extreme value of $\eta$ is 
chosen to understand what the magnitudes of the largest probable corrections 
would be).  Thus, according to equations (4) and (9), though galaxy A has a true 
disk ellipticity of 0.844 it will be observed to have an ellipticity of 0.755 
($i$ = 77.9\deg).  Similarly, galaxy B has \epst\ = 0.324 but \epso\ = 0.290 ($i$ = 45.2\deg).  We see that without accounting for the circularizing effect 
atmospheric seeing has on these disks, one would incorrectly infer the velocity 
widths for galaxies A and B to be 306 \kms\ and 314 \kms\ instead of the actual 
300 \kms.

Seeing also affects the correction for internal extinction.  According to 
equation (1), galaxy A ``should'' have a correction of 0.742 mag and galaxy B, 
0.156 mag, but the observed axial ratios would imply corrections of only 0.562 
and 0.137 mag.

This simple example highlights how seeing affects differently oriented disks.  
Without making the proper corrections for seeing, galaxies that are highly 
inclined would yield reasonable velocity widths but the internal extinction 
corrections for the galaxies would be uncomfortably awry.  Conversely, galaxies 
considered to be borderline TF candidates due to their low inclinations are 
not affected greatly by miscalculating the internal extinction, but errors in 
determining the inclinations would produce velocity width errors of \about\ 5\%. 

\section {Data}					

We present here the relevant data for the first set of seven Abell clusters,
including \Ntot\ galaxies for which we have obtained both photometric parameters
and rotational velocity widths.  We classify \Nuse\ of these galaxies to be 
cluster members with good widths; 15 galaxies are foreground or background 
objects or are cluster members with poorly defined velocity widths.  

Table 1 lists the main parameters of the clusters.  Standard names are listed in 
column 1.  Adopted coordinates of the cluster center are listed in columns 2 and 
3, for the epoch 1950; they are obtained from ACO, except for the entry A1983b, 
a system found to be slightly offset from A1983 in both sky position and 
redshift.  For all the clusters we derived a new systemic velocity, combining the redshift measurements available in the NED\footnote{The NASA/IPAC Extragalactic Database is operated by the Jet Propulsion Laboratory, California Institute of 
Technology, under contract with the National Aeronautics and Space 
Administration.} database with our own measurements.  These newly determined 
velocities are listed in columns 4 and 5, in the heliocentric and in the CMB 
(Kogut \etal\ 1993) reference frame, respectively.  We list the number of cluster member redshifts used in determining systemic velocities in column 6.  An 
estimated error for the systemic velocity is parenthesized after the heliocentric figure.  Spherical and Cartesian supergalactic coordinates are given in columns 7 and 8, and in columns 9--11, respectively.

Figures 3--5 show the distribution of the galaxies in each cluster.  The top 
panel in each of these figures displays the spatial location of:  the outline of
the fields imaged (large squares), cluster members (circles -- those with 
poor/unusable velocity widths are left unfilled), background or foreground 
objects (asterisks), and galaxies with known redshift but without reliable 
widths (dots).  Circles of 1 and 2 Abell radii, \rA, are drawn as dashed 
lines, if the area displayed is large enough.  If no dashed circle is drawn, 
\rA\ is larger than the figure limits.  We also plot radial (CMB) velocity as a
function of angular distance from the cluster center in the lower panel of each
figure.  A dashed vertical line is drawn at 1 \rA.  The combination of the sky
and velocity plots is used to gauge cluster membership for each galaxy.  We 
point out that each of the filled circles do indeed identify {\it bona fide} 
cluster members -- all but two of the galaxies we label as cluster members lie 
within 1 \rA\ and all cluster members are within 1300 \kms\ of the cluster 
systemic radial velocity.

We separate photometric data and spectroscopic data in two tables.  Table 2 lists the spectroscopic properties and table 3 gives the pertinent photometric results.

Entries in the tables are sorted first by the Right Ascension of each cluster, 
and within each cluster sample by increasing galaxy Right Ascension. The 
listed parameters for table 2 are:

\noindent 
Col 1: identification names corresponding to a coding number in our private
database, referred to as the Arecibo General Catalog, which we maintain for easy 
reference in case readers may request additional information on the object.

\noindent
Cols. 2 and 3: Right Ascension and Declination in the 1950.0 epoch.  Coordinates
have been obtained from the Digitized Sky Survey catalog and are accurate to
$<$ 2\arcsec.

\noindent
Cols. 4 and 5: the galaxy radial velocity as measured in the heliocentric and
CMB reference frame (Kogut \etal\ 1993).  Errors are parenthesized: \eg\ 
13241(08) implies 13241$\pm$08.

\noindent
Col. 6: the raw velocity width in \kms. Measurement of optical widths are 
described in section 4; 21 cm line widths are denoted with a dagger and refer to 
values measured at a level of 50\% of the profile horns.

\noindent
Col. 7: the velocity width in \kms\ after correcting for ORC shape and, for 21 cm
data, signal to noise effects, insterstellar medium turbulence, and instrumental and data processing broadening; details on the adopted corrections for optical
and 21 cm data are given in section 4 and G97a, respectively.

\noindent
Col. 8:  the corrected velocity width converted to edge--on viewing, in \kms,
after accounting for the cosmological stretch of the data.

\noindent
Col. 9: the adopted inclination $i$ of the plane of the disk to the line of 
sight, in degrees, (90$^\circ$ corresponding to edge--on perspective); the 
derivation of $i$ and its associated uncertainty are discussed in section 4.

\noindent
Col. 10: the logarithm in base 10 of the corrected velocity width (value in 
column 7), together with its estimated uncertainty between brackets. The 
uncertainty takes into account both measurement errors and uncertainties arising 
from the corrections. The format 2.576(22), for example, is equivalent to 
2.576$\pm$0.022.

The position angle adopted for the slit of each spectroscopic observation is 
that given in column 4 of table 3.  The first column in table 3 matches that of 
table 2.  The remaining listed parameters for table 3 are:

\noindent
Col. 2: morphological type code in the RC3 scheme, where code 1 corresponds 
to Sa's, code 3 to Sb's, code 5 to Sc's and so on.  When the type code is 
followed by a ``B'', the galaxy disk has an identifiable bar.  We assign these
codes after visually inspecting the CCD I band images and after noting the value
of $R_{\rm 75}/R_{\rm 25}$, where $R_X$ is the radius containing X\% of the I band flux.
This ratio is a measure of the central concentration of the flux which was
computed for a variety of bulge--to--disk ratios.  Given the limited resolution
of the images, some of the inferred types are rather uncertain.

\noindent
Col. 3: the angular distance $\theta$ in arcminutes from the center of each 
cluster.

\noindent
Col. 4: position angle of the major axis of the image, also used for 
spectrograph slit positioning (North: 0$^{\circ}$, East: 90$^{\circ}$).

\noindent
Col. 5: observed ellipticity of the disk.
 
\noindent
Col. 6: ellipticity corrected for seeing effects as described in section 5,
along with its corresponding uncertainty.  The format 0.780(16), for example,
is equivalent to 0.780$\pm$0.016.

\noindent
Col. 7: surface brightness at zero radius, as extrapolated from the fit to the
disk surface brightness profile.

\noindent
Col. 8: the (exponential) disk scale length.

\noindent
Col. 9: the distance along the major axis to the isophote containing 83\% of the 
I band flux.

\noindent
Col. 10: isophotal radius along the major axis where the surface brightness 
equals 23.5 mag sec$^{-2}$.

\noindent
Col. 11: apparent magnitude within the 23.5 mag sec$^{-2}$ isophote.

\noindent
Col. 12: the measured I band magnitude, extrapolated to infinity assuming 
that the surface brightness profile of the disk is well described by an 
exponential function.

\noindent
Col. 13: the apparent magnitude, to which k--term, galactic and internal
extinction corrections were applied; details on the adopted corrections are 
given in section 3. 

\noindent
Col. 14: the absolute magnitude, computed assuming that the galaxy is at the 
distance indicated either by the cluster redshift, if the galaxy is a cluster 
member, or by the galaxy redshift if not.  The calculation assumes 
$H_\circ = 100h$ \kms Mpc$^{-1}$, so the value listed is strictly 
$M_{\rm cor} - 5\log h$.  In calculating this parameter, radial velocities are  
expressed in the CMB frame and uncorrected for any cluster peculiar motion. 
The uncertainty on the magnitude, indicated between brackets in hundredths
of a mag, is the sum in quadrature of the measurement errors and the estimate of 
the uncertainty in the corrections applied to the measured parameter.

When an asterisk appears at the end of the record, a  detailed comment is given 
for that particular object.  Because of the length and number of these comments, 
they are not appended to the table but included in the text as follows.  Note
that a record is flagged in both tables 2 and 3, independently on whether the
comments refer only to the photometry, only to the spectroscopy, or both.

\noindent {\bf A168:}\\
\small
\noindent 110868: HII distribution patchy and does not extend far radially; 
exponential disk over narrow range of radii.\\
\noindent 110877: asymm. ORC.\\
\noindent 110879: asymm. central region, both in image and ORC; difficult to 
determine ellipticity.\\
\noindent 410553: mostly bulge ORC; tiny gal.

\normalsize
\noindent {\bf A397:}\\
\small
\noindent 121137: background gal.\\
\noindent 120587: prob. used incorrect PA; 21 cm width preferred; note low i.\\
\noindent 121055: background gal.\\
\noindent 121143: tiny gal; ORC still rising at optical radius; poorly 
determined 21 cm width.\\
\noindent 121145: background gal.\\
\noindent 121060: background gal.\\
\noindent 121061: background gal.\\
\noindent   2406: patchy HII; good agreement between 21 cm and ORC width; used
N[II] ORC.\\
\noindent 120614: only 21 cm width available; note early type; unfit for TF
use.\\
\noindent 121154: mostly bulge ORC; uncertain vel. width.\\
\noindent 121157: background gal.\\
\noindent 120632: poor HII sampling; 21 cm width available\\
\noindent   2420: asymm. HII distrib.; compatible ORC and 21 cm widths.\\
\noindent 120651: asymmetric ORC; PA uncert.; 21 cm width used.\\
\noindent 120652: \hal\ absorption in bulge; note low i; good agreement between
21 cm and ORC width.\\
\noindent 121080: note low i.; compatible 21 cm and ORC widths.\\
\noindent 121082: only 21 cm width available; caution: nearby star may perturb 
phot.\\
\noindent 121085: patchy HII; uncertain PA.\\
\noindent   2453: uncertain ellipticity; \hal\ absorption in bulge; W$_{\rm ORC}$ = 360 \kms; 21 cm width available.

\normalsize
\noindent {\bf A569:}\\
\small
\noindent 170433: ORC still rising.\\
\noindent 170437: background gal.\\
\noindent 170438: ORC only from bulge.\\
\noindent 170439: ORC still rising?\\
\noindent 140440: note low i.\\
\noindent 170442: ORC slightly rising.\\
\noindent 170446: exponential disk over small range of radii.\\
\noindent 170057: ORC slightly rising.\\
\noindent   3724: only 21 cm width available.\\
\noindent 170063: Sd or irregular morphology?\\
\noindent 170468: background gal.

\normalsize
\noindent {\bf A1139:}\\
\small
\noindent 200723: mostly bulge ORC.\\
\noindent 200743: mostly bulge ORC.\\
\noindent 201069: mostly bulge ORC; uncert. vel. width.
 
\normalsize
\noindent {\bf A1228:}\\
\small
\noindent   6357: only 21 cm width available.\\
\noindent   6364: dust lane along major axis; opt. radius signif. larger than 
ORC's extent; used N[II] ORC.\\
\noindent 211475: background gal.\\
\noindent 211476: bar distorts ORC; unfit for TF use.\\
\noindent 211406: asymm. ORC; tiny disk.

\normalsize
\noindent {\bf A1983:}\\
\small
\noindent 241306: only 21 cm width available.\\
\noindent 240823: only 21 cm width availablle.\\
\noindent 240840: mostly bulge ORC.\\
\noindent 241298: weak, patchy HII distrib, but ORC fine.\\

\normalsize
\noindent {\bf A1983b:}\\
\small
\noindent 241056: background gal.; not used in fits.\\
\noindent   9538: unreliable ORC; severe \hal\ absorption; 21 cm width 
available.\\
\noindent 241059: only 21 cm width available.\\
\noindent 241061: background gal.; \hal\ partially absorbed and $\lambda$6548
completely absorbed!.\\
\noindent 241296: large extrapolation to ORC.\\
\noindent 240746: only 21 cm width available.\\
\noindent 240747: 21 cm width available; interacting w/faint companion? prob.
unfit for TF use.\\
\noindent   9548: low vel. horn in 21 cm data poorly defined.\\
\noindent   9550: compatible ORC and 21 cm widths.
\normalsize

In figure 6, we plot the ORCs folded about a kinematic center as described in 
section 4.  The horizontal dashed line in each panel indicates the adopted (and 
uncorrected for inclination) half velocity width, $W$/2, for each galaxy and the 
vertical dashed line is drawn at \ropt.  Finally, in figure 7 we give, as an 
example, the surface brightness profiles for the first 16 galaxies in figure 6.  
Again the vertical dashed line refers to \ropt.  The solid line drawn along 
the disk is the fit to the disk over the range of radii assumed to cover the 
exponential portion of the disk.  The remainder of the plots of surface 
brightness profiles for the complete sample can be obtained by contacting the
first author.

Figure 8 gives the ``raw'' TF plots of each cluster uncorrected for any cluster
incompleteness bias.  A computation of such bias will be presented in future 
work when data from all clusters is in hand.  Furthermore, the cluster systemic
redshifts used in obtaining these plots are preliminary.  A given cluster's
systemic velocity is inferred from all available redshifts in its field.  While
we have tried to target clusters with a broad redshift database (or for which 
redshift surveys are ongoing, as we have learned from informal contacts with 
colleagues), the definition of a cluster's $<\!\!{\rm c}z\!\!>$ will be of 
variable quality.  The peculiar velocity inferred from the TF relation will be 
affected by this uncertainty.  The exact amplitude of that uncertainty will be 
estimated when all the redshift data will become available, at the completion of 
this study.  {\it Thus the TF relations presented in figure 8 are to be 
considered preliminary}.  Included in the TF plots is the template relation 
obtained from nearby clusters in G97b:
\be
y = -7.68x - 21.01
\ee
where $y$ is $M_{\rm cor}$ -- 5log$h$ and $x$ is log$W_{\rm cor}$ -- 2.5.

\acknowledgements

We thank Katie Jore for the use of her ORC fitting programs.
The results presented here are based on observations carried out at the Palomar
Observatory (PO), at the Kitt Peak National Observatory (KPNO), and the Arecibo 
Observatory, which is part of the National Astronomy and Ionosphere Center 
(NAIC).  KPNO is operated by Associated Universities for Research in Astronomy 
and NAIC is operated by Cornell University, both under cooperative agreements 
with the National Science Foundation.  The Hale telescope at the PO is operated 
by the California Institute of Technology under a cooperative agreement with 
Cornell University and the Jet Propulsion Laboratory.  This research was 
supported by NSF grants AST94-20505 to RG and AST90-14850 and AST90-23450 to MH.

\newpage

\figcaption{A comparison of folding techniques.  The upper panel 
gives an ORC folded about the galactic {\it center--of--light} whereas the lower 
panel displays the ORC folded about the {\it kinematic} center of the galaxy.  
The redshift of the galaxy changes by only 34 \kms\ and the radii are shifted by 
only 0.92\arcsec, but the effect is striking.  We adopt the lower panel 
convention. \label{fig1}}

\figcaption{Results of simulations on the effects of seeing on disk ellipticities.  
$\eta$ is the ratio of the seeing FWHM to \rd.  Plotted are the 
data points for simulations carried out on 11 galaxies of differing 
bulge--to--disk ratios and ellipticities.  The ``true'' ellipticities (as 
measured from the raw, unconvolved images) of these 11 galaxies are given in the 
lower right.  We use the linear relation drawn to recover ``true'' 
ellipticities. \label{fig2}}

\figcaption{Sky and velocity distribution of galaxies in the 
clusters Abell 168 and Abell 397.  Circles represent cluster members with measured 
photometry and widths; if unfilled, widths are poorly determined.  Asterisks 
identify foreground and background galaxies and dots give the location of 
galaxies with known redshift, but lacking accurate width and/or photometry.  
Large square boxes indicate outlines of imaged fields with the 0.9 m telescope. 
Vertical dashed lines in the lower panels indicate 1 \rA.  The upper panels 
contain circles of radius 1 \rA\ (and 2 \rA\ for A168). \label{fig3}}

\figcaption{Sky and velocity distribution of galaxies in the 
clusters Abell 569 and Abell 1139.  Filled circles, unfilled circles, asterisks, 
dots, large squares and dashed lines and circles follow the same convention as in
Figure 3.  Note that the upper panel for Abell 569 is smaller than a circle of 
radius 1 \rA. \label{fig4}}

\figcaption{Sky and velocity distribution of galaxies in the 
clusters Abell 1228, Abell 1983, and Abell 1983b.  For A1228 and A1983, filled 
circles, unfilled circles, asterisks, dots, large squares and dashed lines and 
circles follow the same convention as in Figure 3.  For A1983b, filled triangles 
identify cluster members and \rA\ is indicated by the dotted line in the lower panel.
\label{fig5}} 

\figcaption{\hal\ rotation curves for \Nopt\ galaxies (except for 
galaxies 6364 and 2406, for which the ORC is obtained from a N[II] emission 
line), folded about the kinematic centers (see figure 1).  The error bars include
both the uncertainty in the wavelength calibration and the ORC fitting routine 
used.  Names of the galaxy and the corresponding parent cluster are given along 
with the CMB radial velocity.  Two dashed lines are drawn: the horizontal line 
indicates the adopted half velocity width, W/2, which in some cases arises from 
an extrapolation to the ORC or from a 21 cm width (see Table 2); the vertical 
line is at \ropt, the radius containing 83\% of the I band flux.  Note that 
the rotation curves are {\it not} deprojected to an edge--on orientation. 
\label{fig6}}

\figcaption{A sampling of surface brightness profiles.  
Names of the galaxy and the corresponding parent cluster are given in each panel.
 Two lines are drawn: the vertical dashed line is drawn at \ropt\ and the solid 
line is an exponential fit to the disk, over the range of radii over which the 
disk is assumed to behave exponentially. \label{fig7}}

\figcaption{``Raw'' TF plots for the seven clusters are given.  We 
emphasize that the data have {\it not} been corrected for incompleteness bias.  
We also note that A397 is seen through a region of the Milky Way with a large and 
unevenly distributed extinction which leads to increased scatter.  In the A1983 
panel, the error bars containing filled circles represent members of ``A1983b,'' 
a cluster at a slightly lower redshift than A1983.  The dashed line is the 
template relation valid for low $z$ clusters, eqn. (10). \label{fig8}}

\end{document}